\documentclass[twocolumn]{aastex631}

\usepackage{amsmath}
\usepackage{multirow}

\begin{document}

\title{Beyond UV: Rest-frame B-band and Apparent Luminosity Functions of $z=5-9$ Galaxies}

\correspondingauthor{Nicha Leethochawalit}
\email{nicha@narit.or.th}

\author[0000-0003-4570-3159]{Nicha Leethochawalit}
\affiliation{National Astronomical Research Institute of Thailand (NARIT), Mae Rim, Chiang Mai, 50180, Thailand}

\author[0000-0002-8512-1404]{Takahiro Morishita}
\affiliation{IPAC, California Institute of Technology, MC 314-6, 1200 E. California Boulevard, Pasadena, CA 91125, USA}

\author[0000-0002-0384-305X]{Tirawut Worrakitpoonpon}
\affiliation{School of Physics, Institute of Science, Suranaree University of Technology, Nakhon Ratchasima 30000, Thailand}

\author[0000-0001-9391-305X]{Michele Trenti}
\affiliation{School of Physics, University of Melbourne, Parkville 3010, VIC, Australia}

\begin{abstract}
We present new measurements of galaxy luminosity functions (LFs) from JWST/NIRCam imaging over the redshift range $z=4.5–9.7$, using photometric catalogs from JADES and public extragalactic fields. Our analysis includes rest-frame UV and B-band LFs, as well as apparent LFs in F090W, F115W, F200W, F356W, and F444W. We present the first constraints on the rest-frame B-band LF at $z\sim7$–8 and extend existing measurements at $z\sim5$ to $M_B = -18$ mag. The B-band LFs evolve more strongly with redshift than UV LFs, though both decline more gradually than predicted by simulations at $z>5$. No single existing simulation reproduces all observed trends, with discrepancies likely driven by assumptions about binary evolution and stellar population synthesis models. The apparent LFs in F356W and F444W show hints of a bright-end excess at all redshifts, extending to fainter magnitudes at higher redshift. While extreme emission line galaxies may partially account for it, the excess may also indicate a population of moderately red, optically bright sources—potentially dusty star-forming galaxies or obscured AGNs. Finally, we find that rest-frame B-band luminosity correlates more tightly with stellar mass than UV, making it a powerful tracer of mass assembly and reinforcing the diagnostic value of rest-frame optical LFs in uncovering the physical processes that drive early galaxy formation.

\end{abstract}



\section{Introduction} \label{sec:intro}

The ultraviolet (UV) luminosity function (LF) has long been a cornerstone of high-redshift galaxy studies. Initially, it was used to trace the co-moving star formation rate density over cosmic time \citep[e.g.,][]{Lilly1996,Reddy2009}, and later to constrain the ionizing photon budget and assess the role of galaxies in cosmic reionization \citep[e.g.,][]{Robertson2015,Finkelstein2019}. However, due to its observational availability, the UV LF has become the main benchmark for testing galaxy formation models at early times. Simulations are now routinely calibrated to match the UV LFs at $z>5$ \citep[e.g.,][]{Sipple2024,Donnan2025}.

Yet, there is no fundamental reason to limit observational constraints to the rest-frame UV. The shape and evolution of the luminosity function in other bands—such as the rest-frame optical—can offer complementary insights. While UV light traces young, massive stars and recent star formation, optical light originates from longer-lived stars and is therefore more sensitive to the integrated star formation history. Observing luminosity functions across multiple rest-frame wavelengths can help break degeneracies in spectral energy distribution (SED) used in simulations, particularly those involving the stellar population, and dust content.

Recent James Webb Space Telescope (JWST) observations have revealed a surprising abundance of bright galaxies at $z>7$, especially in the rest-frame UV. The bright end of the UV LF appears to evolve only mildly at $z\gtrsim7$ \citep[e.g.,][]{Donnan2024, Robertson2024, Rojas-Ruiz2024, Morishita2025}, confirming earlier results from HST parallel surveys \citep{Leethochawalit2023, Bagley2024}. However, the underlying physical nature of these UV-bright galaxies remains uncertain. 

Several mechanisms have been proposed to explain the high number densities of UV bright galaxies at early times without invoking changes to cosmology. These include enhanced star formation efficiency \citep{Mason2023, Harikane2023, Li2024}, scatter in the stellar-to-halo mass relation \citep{Gelli2024}, a top-heavy IMF \citep{ Ventura2024}, and UV contributions from active galactic nuclei (AGNs) \citep{Hegde2024,Trinca2024}. Importantly, these scenarios may predict different outcomes in redder bands. For instance, a top-heavy IMF and dust obscuration affects rest-frame UV more than rest-frame optical light. Thus, rest-frame optical luminosity functions offer a promising avenue for disentangling these physical drivers.

Historically, rest-optical LFs have been studied at lower redshifts ($z\lesssim5$) in the context of the evolution of blue and red galaxy populations \citep[e.g.,][]{Poli2003, Gabasch2004, Giallongo2005}. These studies found that the redshift evolution is more pronounced at bluer wavelengths and that faint-end slopes are generally steeper in the $g'$ and B bands than in the UV \citep[e.g.,][]{Gabasch2004}. For the wavelength, the B-band luminosity is less susceptible to bursty star formation activities and dust attenuation, enabling less biased evaluation of galaxy luminosity functions at early cosmic times. This makes the analysis of the B-band LF potentially valuable in the context of the aforementioned recent JWST findings.
However, the rest-frame B-band LF has not been reliably constrained beyond $z\sim5$ due to the limited wavelength coverage of HST and ground-based facilities.

In parallel, theoretical works have begun to predict apparent magnitude functions, particularly in JWST/NIRCam bands \citep{Yung2019, Vogelsberger2020}, as a direct way to compare simulations with forthcoming observations. These functions—defined as the number of galaxies per apparent magnitude per comoving volume—avoid assumptions about distance modulus or SED fitting, making them less sensitive to redshift uncertainties. Although their physical interpretation is less straightforward than that of rest-frame LFs, they provide a valuable bridge between simulations and observational datasets.

The advent of JWST has made it possible to obtain the first robust measurements of rest-frame optical and apparent JWST-band luminosity functions at $z>6$. At these redshifts, the rest-frame B band is shifted into the near- and mid-infrared, where NIRCam’s broad wavelength coverage provides the necessary sensitivity. By measuring luminosity functions directly in multiple NIRCam filters, we can analyze large galaxy samples based solely on photometry, offering a complementary perspective on galaxy evolution that minimizes dependence on photometric redshifts and SED-fitting assumptions.

In this paper, we present new measurements of galaxy luminosity functions at $z\sim5$--9 in rest-frame UV, rest-frame B, and multiple observed NIRCam bands (F090W, F115W, F200W, F356W, and F444W). We use deep photometric catalogs from the JADES and other public JWST extragalactic fields and apply consistent selection criteria and completeness corrections based on injection-recovery simulations. Our results offer the first constraints on the rest-frame B-band LF at $z\sim7$ and 8, providing a new empirical benchmark for models of early galaxy formation and stellar population synthesis.

\section{Data} \label{sec:data}
The most common reddest filter in typical NIRCam observations is F444W which corresponds to rest-frame B band ($\lambda_\textrm{eff}\sim 4300$\AA) at $z\sim9$. Similarly, the most common bluest filter is F090W which corresponds to rest-frame UV ($\lambda=150$ nm) at $z\sim4.5$. We are therefore interested in the redshift sample of $z\sim5$ to 9. 

In this paper, we get the data sample from two datasets. For $z\sim5-7$, we use the v.2.0 photometric catalogs from the JWST Advanced Deep Extragalactic Survey \citep[][JADES]{Rieke2023}. For the $7.2<z<9.7$ (F090W dropout) sample, we require a larger area to get a sufficient number of galaxies. Thus, we use the public deep NIRCam fields from JWST Cycle 1 that have been homogeneously compiled and reduced by \citet{Morishita2024}. 

\subsection{Data Selection of $z\sim5-7$ galaxies}\label{sec:lowz-selection}
To reliably select $z\sim5-7$ galaxies, fields with existing deep optical observations are necessary. Here, we use the JADES second data release of the GOODS-S vicinity \citep{Eisenstein2023} which handily provides fluxes in the optical HST bands that are PSF-matched to the JWST images. In this catalog, the objects were detected from a stacked signal-to-noise image created from the F277W, F335M, F356W, F410M, and F444W data \citep{Rieke2023}. Based on the catalog, we use the following selection criteria.

\begin{flalign*}
    \textrm{F435W dropouts}:&&\\ 
    &\textrm{SNR}_\textrm{F090W}>5 \\
    &\textrm{SNR}_\textrm{F435W}<2\\
    &4.5\leq z_p < 6\\
    &P(z_p > 4.5) > 0.84
\end{flalign*}
\begin{flalign*}
    \textrm{F606W dropouts:}&&\\
    &\textrm{SNR}_\textrm{F115W}>5 \\
    &\textrm{SNR}_\textrm{F435W,F606W}<2\\
    &6.0 \leq z_p < 7.5\\
    &P(z_p > 6.0) > 0.84
\end{flalign*}
The signal-to-noise ratios (SNRs) are calculated from circular apertures with 0.15\arcsec radius. These dropout criteria essentially require detections with SNR of at least 5 in the rest-frame UV band and no detection in the bands bluer than the Lyman break to ensure the dropout nature and reduce contamination from red low-z galaxies. We use the photometric redshifts provided in the catalogs, which were calculated with the \texttt{eazy} photometric redshift code \citep{Brammer2008} and the galaxy templates from \citet{Hainline2024}.

To constrain the LFs in rest-frame UV and B bands, as well as in the observed NIRCam bands, we additionally require the SNR of at least 8 in the specific band used for the LF measurement. This ensures that the objects are directly detected and have well-defined fluxes in the relevant band. For example, F435W (F606W) dropouts must have $\textrm{SNR}_\textrm{F090W} >8$ ($\textrm{SNR}_\textrm{F115W}>8$) to be included in the rest-frame UV LF calculation. Similarly, they must have $\textrm{SNR}_\textrm{F277W}>8$ ($\textrm{SNR}_\textrm{F356W}>8$) to be included in the rest-frame B-band LF calculation. For the observed NIRCam-band LFs --- specifically F090W, F115W, F200W, F356W, and F444W --- we require $\textrm{SNR}>8$ in the respective band. As a result, the number of candidates used to construct the different LFs vary, even though they are of the same dropout band. 

We apply additional criteria to exclude potential spurious sources. Specifically, using the JADES catalog flags, we exclude candidates flagged as stars, affected by bright stars, or neighboring objects more than twice as bright. Visual inspection further reveals some candidates embedded in bright regions, which may affect flux measurements. To minimize this, we exclude candidates with over 50\% of their boundary adjacent to other objects, based on the segmentation map.

To reduce the contamination from ultra cool dwarfs in compact candidates, we verify that their SEDs are better fit with galaxy templates than stellar template. If a candidate is compact, i.e., having a half-light radius in the F444W filter smaller than the PSF FWHM/2 (0.\arcsec073), we refit their photometry with \texttt{eazy} using the SpeX Prism Spectral Libraries \citep{Burgasser2014}. If the resulting $\chi^2$ value is smaller than that from the galaxy template, we remove the candidate. 

Based on these criteria, there are 1380 F435W-dropout and 696 F606W-dropout candidates. The specific numbers of galaxies used for each LF band along with the redshift ranges are listed in Table \ref{tab:numbercandidates}.

\begin{table*}[]
\centering
\begin{tabular}{|c|cccccccc|}
\hline
\multirow{2}{*}{$z$} & \multicolumn{8}{c|}{Number of candidates (Number of expected low-z contaminants)/Total area in arcmin$^2$ for each LF band} \\ \cline{2-9}
                  & UV & B & F090W & F115W & F200W & F356W & F444W & All\\ \hline
$5.2$    & 982(8)/64.2 & 1283(22)/63.5 & 982(7)/64.2 & 1113(12)/64.1 & 954(8)/63.2 & 1317(18)/63.0 & 1058(5)/64.2 & 1380 \\
$6.7$   & 475(3)/64.2 & 631(6)/63.0 & 118(0.2)/64.1 & 475(3)/64.2 & 359(2)/63.2 & 631(6)/63.0 & 453(2)/64.2 & 696  \\
$7.8$    & 240(11)/289.3  & 295(11)/285.9 & -- & 155(2)/277.2 & 244(21)/282.6 & 332(32)/277.5  & 295(11)/285.9 & 406  \\ \hline
\end{tabular}
\caption{Each cell lists the number of candidates, followed in parentheses by the expected number of low-$z$ contaminants, and the total survey area (in arcmin$^2$) used to construct the luminosity function for each band and dropout sample. From top to bottom, the rows correspond to F435W dropouts ($z=4.8$–$5.7$), F606W dropouts ($z=6.4$–$7.1$), and F090W dropouts ($z=7.3$–$8.9$).}
\label{tab:numbercandidates}
\end{table*}

\subsection{Data Selection of $z\sim8$ galaxies}\label{sec:highz-selection}
From all the JWST NIRCam Extragalactic fields compiled in \citet{Morishita2024}, we select the F090W dropouts from the following fields: 
PRIMER-UDS \citep[PID 1837, PI Dunlop,][]{Donnan2024},
J1235 (PID 1063, PI Sunnquist), 
North Ecliptic Pole Time-domain field \citep[PID 2738, PI Windhorst,][]{Windhorst2023},
PAR1199 \citep[PID 1199, PI Stiavelli,][]{Stiavelli2023}, 
JADES-GDS \citep[PID 1180, PI Eisenstein,][]{Rieke2023}, and CEERS \citep[PID1345, PI Finkelstein,][]{Finkelstein2025}. We exclude the A2744 field \citep[PID 1324, PI Treu,][]{Treu2022} which is centered at a massive galaxy cluster to avoid the complication from lensing magnification. We also exclude the NGDEEP field \citep[PID 2079, PI Finkelstein,][]{Bagley2024} because there is no available non-detection bands for our F090W selections. The catalog is based on images that are PSF-matched to the PSF of F444W. It is created with SExtractor using a stack of F277W, F356W, and F444W images as detection images.

We use similar selection criteria as those in \citet{Morishita2024} and those of lower redshift galaxies above,
\begin{flalign*}
    \textrm{F090W-dropouts:}&&\\
    &\textrm{SNR}_{F150W}>5 \\
    &\textrm{SNR}_\textrm{nondet}<2\\
    &7.2 \leq z_p < 9.7\\
    &P(z_p > 6.0) > 0.84
\end{flalign*}
In addition, we require that each candidate must have available photometry in five or more bands and that the SNR in the LF band of interest is larger than 8, e.g., SNR$_{F150W}>8$ for the galaxies in the UV LF calculation and SNR$_{F444W}>8$ for the galaxies in the B-band LF calculation. The non-detection bands for $\textrm{SNR}_\textrm{nondet}$ vary from field to field: F090W for PRIMER-UDS, NEP, PAR1199, and JADES-GDS; F070W and F090W for J1235; and HST's F814W and F606W for the CEERs fields. The detection SNR is measured in $r=0.\arcsec16$ aperture. Since higher redshift galaxies are fractionally more prone to contaminations from lower redshift red galaxies, the no-detection SNR requirement for F090W dropouts is done in both  $r=0.\arcsec08$ and  $r=0.\arcsec16$ apertures. The photometric redshifts are taken from the catalog, which were also measured with \texttt{eazy} using the same template library from \citet{Hainline2024}. For compact candidates, we discard those with $\chi^2$ values from the best-fit stellar templates smaller than the $\chi^2$ from the best-fit galaxy templates, in the same manner done to the lower redshift candidates. We then visually inspect all galaxies to discard artifacts from stellar spikes and objects that are blended with large foreground galaxies.

Based on the criteria and selection procedures above, there are 406 F090W-dropout candidates. The LF-band specific numbers of candidates are also listed in Table \ref{tab:numbercandidates}.

\subsection{Total magnitudes}
For F090W dropouts, we adopt the total fluxes from the \citet{Morishita2024} catalog, where total fluxes were derived following standard procedures for high-redshift sources. Specifically, total fluxes were calculated by scaling aperture fluxes—measured within a 0.\arcsec16 radius aperture—by a correction factor \( C = f_\mathrm{auto,F444W} / f_\mathrm{aper,F444W} \), which accounts for flux outside the aperture. This correction is assumed to be consistent across all bands. For F435W and F606W dropouts from the JADES catalog, we use their Kron-aperture photometry, which has been shown to agree with the total fluxes reported in the CANDELS multiwavelength catalog \citep{Rieke2023}. We note that the commonly-used parameters to derive Kron or \texttt{auto} fluxes, which are also used here, underestimate total fluxes by $10-20\%$ \citep[e.g.,][]{Merlin2019,Leethochawalit2023}. We will correct for this bias in Section \ref{sec:intrinsic_magnitude}.  

\subsection{Rest-frame Absolute Magnitudes}
\label{sec:absmag_measurement}
Using the photometric catalog described above, we measure the rest-frame absolute magnitudes of the candidates with the \texttt{abs\_mag} function in the \texttt{eazy-py} code \citep{Brammer2008}. This function minimizes dependence on SED models by refitting the photometry, down-weighting observed bands that are further from the desired rest-frame band. We fix the redshifts to the photometric redshift estimates. The rest-frame bands measured are $UV_{1500}$—defined as a top-hat filter 100 \AA\ wide and centered at 1500 \AA—and the Johnson-Cousins B filter, which has a pivot wavelength of 4294 \AA\ and a wavelength range of 1311 \AA. Notably, the B filter encompasses two sets of strong emission lines at both its blue and red ends: the [OII] $\lambda\lambda$3726,3729 doublet and the H$\beta$ + [OIII] $\lambda$5007 complex. They fall within regions of approximately 40\% and 25\% of the filter’s peak transmission, respectively.

We compare the \texttt{eazy}-measured rest-frame magnitudes against the \texttt{Bagpipes} fitting code \citep{Carnall2018} and found good agreements. For the \texttt{Bagpipes} fitting procedure, we use the BPASSv2.2 stellar population model \citep{Stanway2018}, assuming non-parametric star formation histories \citep{Iyer2019} and the \citet{Calzetti2000} dust attenuation curve. We measure the absolute magnitudes from the best-fit spectra sampled from the posterior distribution. The measured rest-frame magnitudes from \texttt{eazy} and \texttt{Bagpipes} form a tight correlation. The medians of the differences in the measured rest-frame magnitudes are consistent with zeros within $1\sigma$ in all dropouts and rest-frame bands. 
The only exception is the rest-frame UV magnitude of the F606W dropouts, where the median difference is $0.07^{+0.17}_{-0.06}$ mags. We conclude that, in all cases and bands, the differences between \texttt{eazy}-measured rest-frame magnitudes and \texttt{Bagpipes}-measured rest-frame magnitudes are less than 0.1 mag and are consistent with zeros within $2\sigma$, suggesting that the measured rest-frame magnitudes are not very sensitive to SED fitting choices. 

\subsection{Large-image processing}
Some of the fields—namely JADES, PRIMER-UDS, and J1235—are large mosaics created from multiple pointings with varying depths. The large size of these mosaiced images introduces two challenges for completeness and contamination calculations in subsequent sections. First, the substantial data volume significantly increases computation time. Second, the inhomogeneous depth across the mosaic can reduce the reliability of completeness and contamination estimates.

\begin{figure*}
    \centering 
    \includegraphics[width=0.5\textwidth]{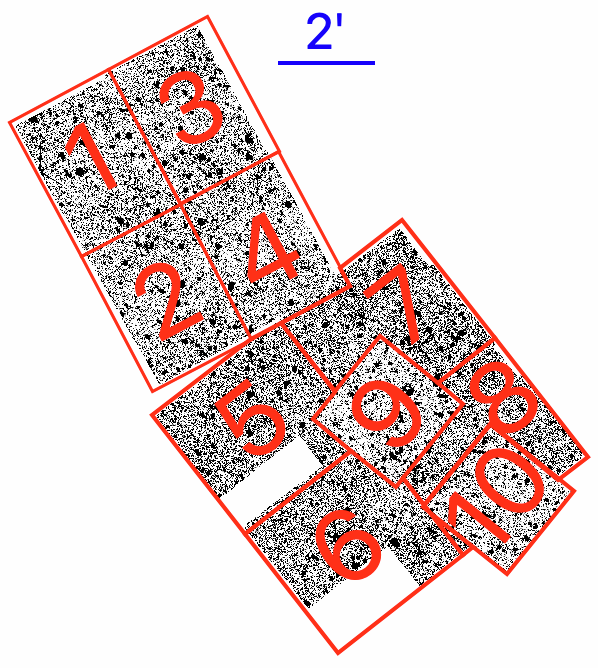}\hspace{1cm}
    \includegraphics[width=0.25\textwidth]{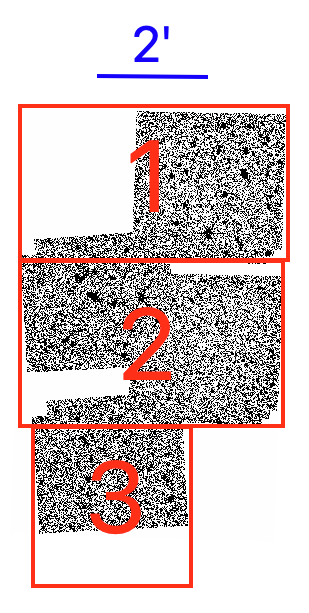}\\
    \includegraphics[width=0.7\textwidth]{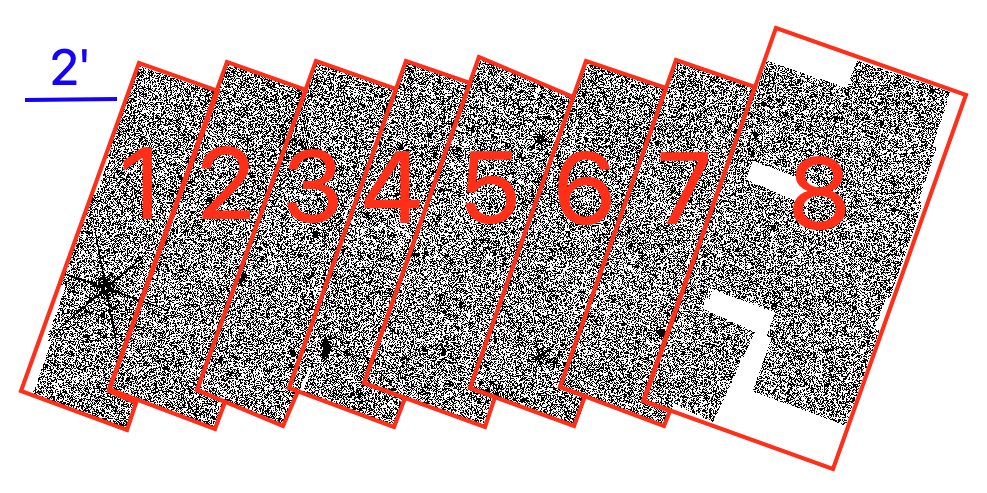}
    \caption{We divide large mosaics into smaller images outlined by red boundaries, in order to better calculate the completeness and contamination by low-z galaxies. The fields are JADES (top left), J1235 (top right), and PRIMER-UDS (bottom).}
    \label{fig:subfields}
\end{figure*}

To mitigate these issues, we divide each large field into smaller subfields of similar depth and trim blank regions outside the observed areas, making the datasets more manageable. The division scheme is illustrated in Figure~\ref{fig:subfields}. These subfields are generated primarily using \texttt{IRAF} \citep{Tody1986} and/or the \texttt{Cutout2D} function in Python. The CEERS fields are already divided into 10 pointings in the original dataset, and we use them as they are.

\subsection{Contaminants from low-z interlopers}\label{sec:contam}
In general, high-redshift galaxy selections may be contaminated by two types of lower-redshift objects: ultra-cool dwarfs (UCDs) and low-redshift galaxies. However, the former is not a major concern for JWST-selected candidates in our redshift range for two reasons. First, the SEDs of UCDs differ significantly from those of high-redshift galaxies once wavelengths redder than 2\,$\mu$m are included \citep[e.g.,][]{Finkelstein2016}. Second, the high spatial resolution of JWST enables reliable morphological discrimination between point sources and extended sources. Since these criteria were already incorporated into our data selection, we assume that contamination from UCDs is negligible. 

Low-redshift contaminants, on the other hand, arise when the Balmer break of a low-redshift galaxy is misidentified as a Lyman break. They are common contaminants in JWST-selected $z>8$ candidates \citep[e.g.,][]{ArrabalHaro2023,Napolitano2025}. To estimate the expected number of such interlopers in our sample, we follow the approach described in \citet{Leethochawalit2023}. These estimates are then used to statistically account for low-redshift contamination in the luminosity function calculation (Section \ref{sec:LFmethod}).

In short, we estimate the number of low-z interlopers by creating mock catalogs of low-z galaxies based on the depth of each filter in the observed (sub)field and then passing them through our galaxy selection criteria. We calculated the depth of each subfield as the $1 \sigma$ standard deviation of the fluxes measured in 500 empty apertures randomly placed in the sky region of each image. The aperture radii are chosen to match the aperture size used in the detection SNR requirement. We then generate mock catalogs of low-z galaxies using all 10 realizations of the JAdes extraGalactic Ultradeep Artificial Relizations (JAGUAR) v1.2 catalogs \citep{Williams2018}, which contain both quiescent and star-forming galaxies corresponding to a random galactic field of 121 arcmin$^2$. Here, we consider galaxies with $0.5 <z < 3$ as potential interlopers. We add noise to the photometry of these galaxies according to the $1\sigma$ depth of each filter in the observed field and then pass these mock catalogs through the selection criteria in Sections~\ref{sec:lowz-selection} and~\ref{sec:highz-selection}. The final expected number of interlopers in each field is taken as the mean numbers of low-z galaxies that pass the selection criteria—scaled to the observed area—across all realizations. 

We list in Table \ref{tab:numbercandidates} the total number of expected low-$z$ interlopers across all fields (shown in parentheses). In general, the expected contamination is below $2\%$ for F435W and F606W dropouts. The fraction increases for F090W dropouts, reaching $5$–$10\%$, consistent with contamination rates found in JWST/NIRSpec follow-up observations of NIRCam/JWST-selected candidates at comparable and higher redshifts \citep{ArrabalHaro2023,Morishita2023,Napolitano2025}.

\section{Methods to Derive Luminosity Functions}\label{sec:LFmethod}

\subsection{Completeness simulation} \label{sec:completeness}
We use an adaptation of the injection-recovery simulation \texttt{GLACiAR2} \citep{Leethochawalit2022} to estimate the effective volumes needed to calculate luminosity functions. Galaxies are injected in bins of redshift and rest-frame $UV_{1500}$ magnitude. Specifically, the redshift bins are $z=[4.0, 8.0]$ for F435W and F606W dropouts, and $z=[6.0,10.0]$ for F090W dropouts, with a bin width of $\Delta z = 0.5$ in both cases. The magnitude bins vary between fields; they are chosen to extend approximately two magnitudes fainter than the faintest candidates in each field, with a minimum range of $[-22, -18]$ mag and a bin width of 0.5 mag. For each bin, we inject roughly one galaxy per 30 arcsec$^2$ of area in the detection image at random positions, i.e., about 1200 galaxies per one NIRCam pointing. The galaxies are of Sersic shape with Sersic index n=1 and have random inclinations and angles. The radii of the injected galaxies follow the $M_\textrm{UV}$-size relation found in \citet{Morishita2024}, which is also a function of redshift. Each galaxy's intrinsic spectrum is randomly drawn from galaxies of the same redshift bins in all 10 realizations of the JAGUAR mock catalog\citep{Williams2018}, renormalized to have the $UV_{1500}$ magnitude equal to the intended injected magnitude. The galaxies are injected into all JWST bands available for each field. They are then recovered with the same procedures and selection method listed in Section \ref{sec:data}. 

The completeness functions for all the photometric bands are based on the above simulation. For the UV band, $P(M_{UV_{1500}},z)$ is the fraction of the number of the recovered to the number of the injected galaxies in that UV magnitude and redshift bin. Every bin has the same number of injected galaxies. 

For other LF bands (i.e., rest-frame B and apparent F090W, F115W, F200W, F356W and F444W bands), we follow the procedure below. First, we compute the intrinsic absolute magnitude in rest-frame B band and determine the apparent magnitude in NIRCam bands based on the injected SEDs. These magnitudes are then grouped into bins of 0.5 mag. The number of injected galaxies is defined as the number of sources within each magnitude bin with injected positions in the ``good" region, i.e., the area covered by all bands necessary for the selection criteria and used for candidate selection of that LF band. This mean that for LF band other than rest-frame UV, the numbers of injected galaxies at each magnitude-redshift bin can vary. However, the variation is typically within 10\%. The completeness in each bin is then calculated as the ratio of the number of recovered galaxies to the number of injected galaxies. The total area of these ``good" region for each dropout and LF band are listed in Table \ref{tab:numbercandidates}. 

\begin{figure*}
    \centering 
    \includegraphics[width=0.32\textwidth]{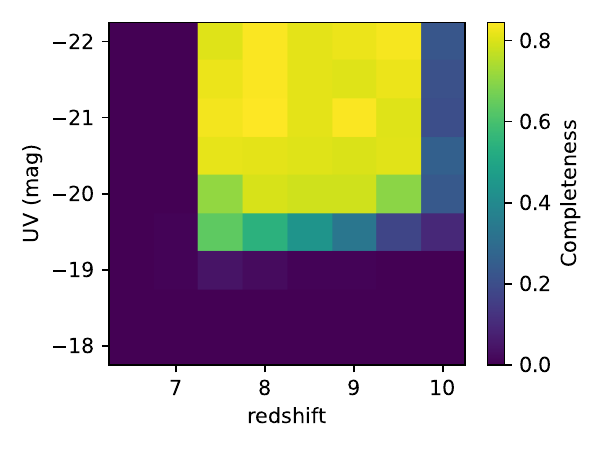}
    \includegraphics[width=0.32\textwidth]{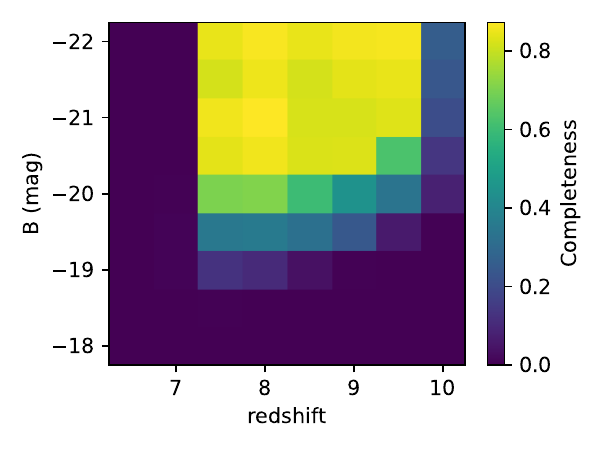}
     \includegraphics[width=0.32\textwidth]{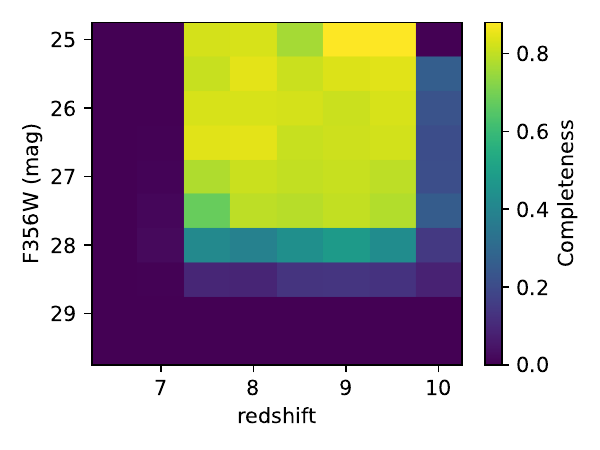}
     \caption{Example of completeness functions for F090W dropouts in the field PAR1199 for UV, B, and F356W band luminosity function constructions (left to right respectively).}
    \label{fig:completeness}
\end{figure*}

The completeness increases with the brightness of the injected sources and plateaus in the brightest magnitude bins. The plateau level of completeness varies across fields, depending on their depths. In the deepest regions—such as the JADES area within the Hubble Ultra Deep Field (areas 9 and 10 in the top-left corner of Figure \ref{fig:subfields})—the completeness plateaus at approximately 65–70\%. In most of the shallower fields, the plateau level is around 80\%. This dependence on depth arises because deeper fields contain a higher density of detected objects. Consequently, injected sources are more likely to overlap with or be blocked by existing objects along the line of sight and are therefore discarded. We show examples of completeness functions for the field PAR1199 in Figure \ref{fig:completeness}.
\subsection{Intrinsic Magnitudes} \label{sec:intrinsic_magnitude}
Various flux estimation procedures can introduce biases to the measured fluxes. We statistically correct the observed magnitudes for these biases during the LF calculation. 
\subsubsection{Intrinsic Apparent Magnitudes}\label{sec:intrinsic_apmag}
From the injection-recovery simulation, we generate a probability distribution function (PDF) of the difference between the injected and recovered apparent magnitudes, binned by recovered magnitude and redshift. We model these PDFs with Gaussian functions, which provide good fits across all bins. In brightest magnitude bins within the highest completeness plateau (e.g. $<27.5$ mag in the JADES fields, and $<26.5$ mag in the shallower J1235 field), the recovered magnitudes are typically fainter than the injected ones by $\mu=0.08-0.12$ mag, corresponding to a $\sim10\%$ flux loss. Near the completeness dropoffs, this bias increases to as much as 0.2 mag. This behavior is consistent with known limitations of SExtractor’s elliptical aperture flux estimation (\texttt{FLUX\_AUTO}). Specifically, when using the commonly adopted minimum Kron radius of $R_\textrm{Kron,min} = 3.5a$, flux losses of approximately $15-20\%$ are expected \citep{Merlin2019}. The flux losses found in our simulations fall within a similar range. 

During the calculation of apparent luminosity functions, we use the Monte Carlo method to correct for these apparent magnitude biases. For each candidate, we sample its magnitude 100 times based on their measured value and uncertainties. For each sampled magnitude, we sample its ``intrinsic'' apparent magnitude based on the previously generated PDF, multiplied by a prior on the number density of the galaxies as a function of magnitude. This prior helps correct for the Eddington bias: it is statistically more likely for a fainter (and more numerous) source to scatter brighter than for a brighter source to scatter fainter. Since there is no existing observational apparent magnitude LFs for our redshift ranges, we adopt simulation-based LFs from \citet{Vogelsberger2020} as our prior. 

We find that applying priors minimally alters the shapes of the PDFs. For the recovered magnitude bins within the highest comleteness plateau, the prior shifts peak locations by $\lesssim0.01$ mag toward the fainter intrinsic magnitudes. This small shift is due to the narrow widths ($\sigma\lesssim0.05$ mag) of the original PDFs in these bright bins. Near the completeness dropoffs, however, the priors have a more noticeable effect; they reduce the discrepancy between the injected and recovered magnitudes from about 0.20 mag (without priors) to 0.12–0.15 mag.

\subsubsection{Intrinsic Absolute Magnitudes}
For rest-frame UV and B magnitude, biases in redshift measurements also contribute to biases in absolute magnitude estimates. Our completeness simulations show that the recovered photometric redshifts from EAzY exhibit biases whose direction and magnitude depend on the injected redshift relative to the target redshift of the selection criteria. At injected redshift bins near the median redshift of each dropout selection, the redshift biases—quantified as $\langle z_p - z_\textrm{injected} \rangle / (1 + z_\textrm{injected})$—are small, ranging from $0.1$ to $-2\%$. The bias is smallest in the JADES field, where more filters are available. At other injected redshifts, however, the direction of the bias varies. For galaxies injected at redshifts lower than the target redshift, the recovered redshifts tend to be biased high. For example, galaxies injected at $z = 7$ into the F090W-dropout selection, the recovered redshifts are biased high, with a median $\langle z_p - z_\textrm{injected} \rangle \sim 0.5$. In contrast, galaxies injected at $z = 9.5$ show a low-redshift bias, with a median $\langle z_p - z_\textrm{injected} \rangle \sim -0.3$. This behavior arises because the F090W-dropout selection requires photometric redshifts in the range $7.2 \leq z_p < 9.7$; only sources that scatter into this window are included. Nevertheless, we do not expect galaxies with true redshifts near or outside the dropout boundaries to contribute significantly to the observed sample, since completeness is low at those redshifts. 

We implicitly include the correction for redshift in the correction for magnitudes. We calculated the PDFs of the intrinsic magnitudes at a given recovered magnitude, using the same method applied to the apparent magnitude. The intrinsic absolute magnitudes are then resampled from the PDFs. With this approach, the bias in redshift estimates is already incorporated into the magnitude correction. For the priors, we adopt the Schechter function parameters from \citet{Bouwens2021} for F435W and F606W dropouts and from \citet{Morishita2025} for the F090W dropouts. 

\begin{figure*}
    \centering 
    \includegraphics[width=0.45\textwidth]{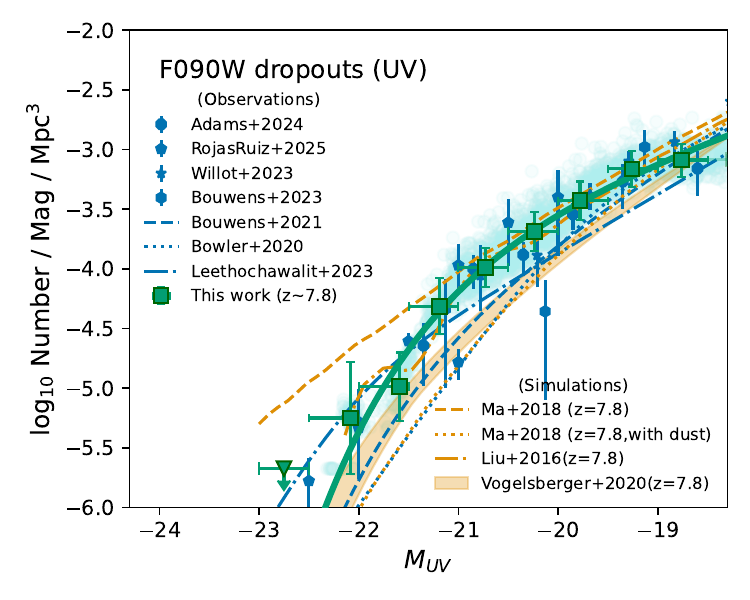}
    \includegraphics[width=0.45\textwidth]{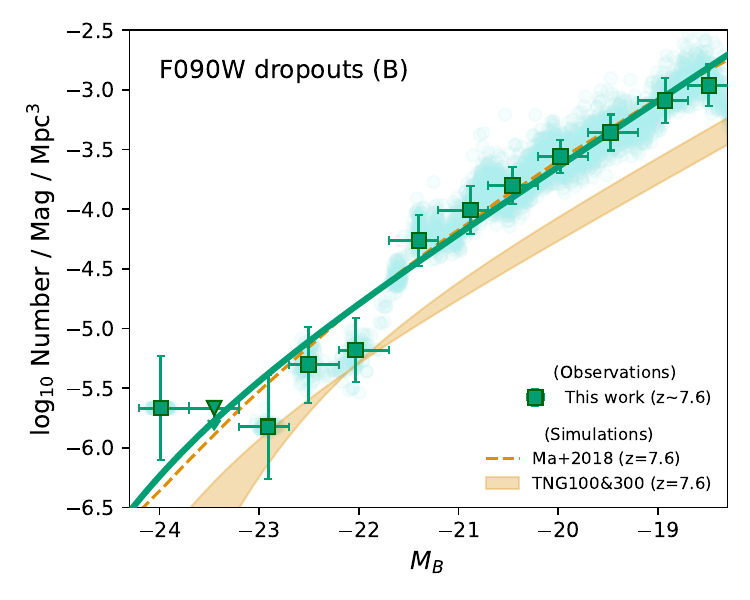}\\
    \includegraphics[width=0.45\textwidth]{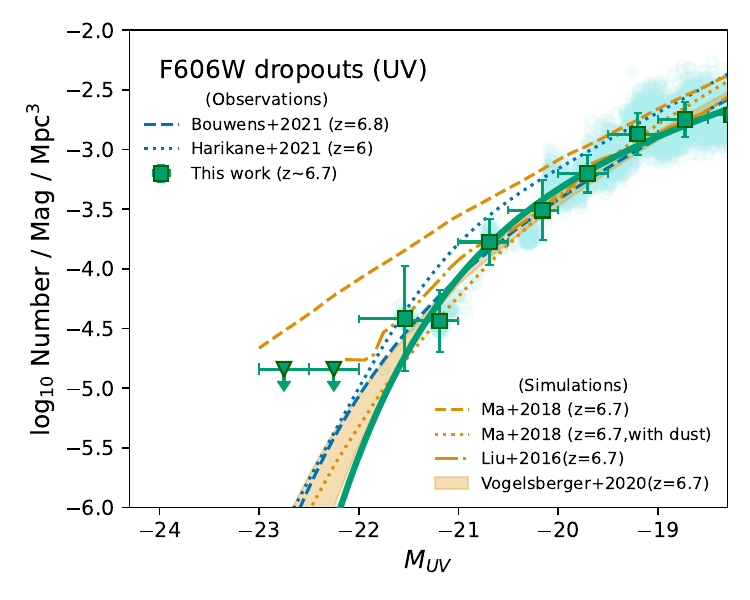}
    \includegraphics[width=0.45\textwidth]{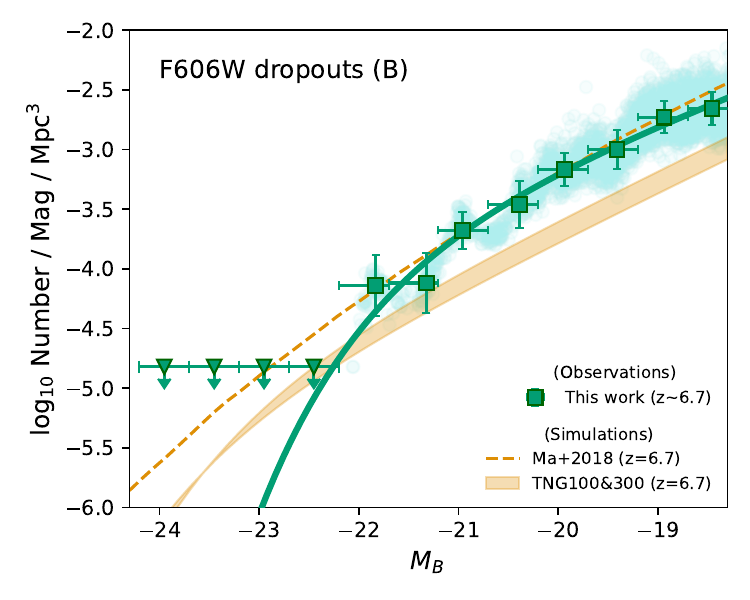}\\
    \includegraphics[width=0.45\textwidth]{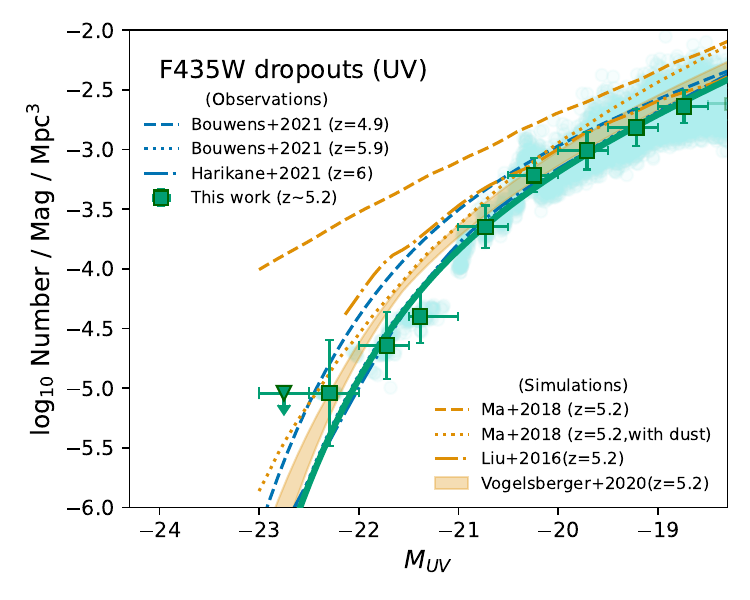}
    \includegraphics[width=0.45\textwidth]{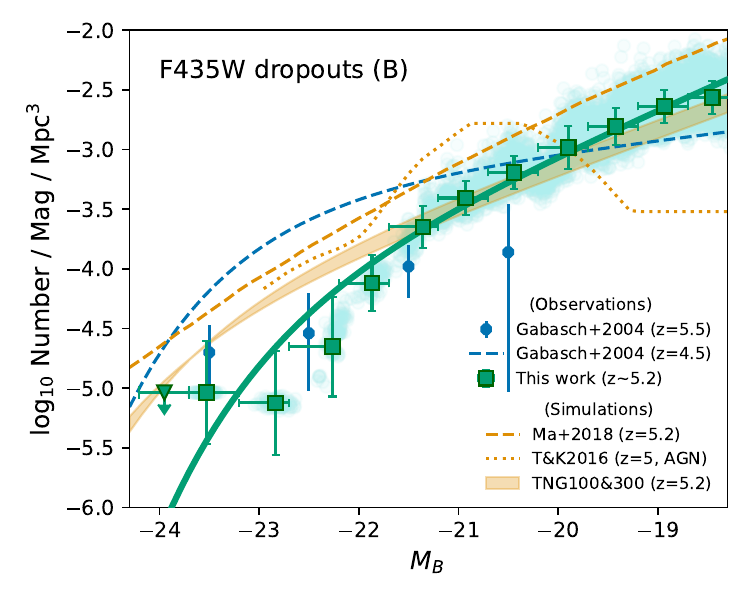}
    \caption{Rest-frame UV and B-band luminosity functions for F090W dropouts (top row), F606W dropouts (middle row), and F435W dropouts (bottom row). Green square data points represent the nominal luminosity functions, while green solid lines show the best-fit Schechter functions. Transparent green points in the background show the individual luminosity functions derived from each Monte Carlo iteration. Observational results from the literature \citep{Bouwens2021, Harikane2022, Giallongo2005, Gabasch2004, Adams2024, Rojas-Ruiz2024, Willott2024, Bouwens2023, Leethochawalit2023} are shown as blue data points and lines. Predictions from cosmological simulations — including those by \citet{Ma2018, Vogelsberger2020} and the TNG series — are shown in brown, linearly interpolated from integer redshifts to the median redshifts of our sample. The nominal luminosity functions and the corresponding best-fit Schechter parameters shown in this figure are provided as the Data behind the Figure.}
    \label{fig:LumfuncUV}
\end{figure*}

\subsection{$1/V_{max}$ method}
\label{sec:vmax}
We first calculate the effective volume as a function of magnitude.
\begin{equation}
    V_\textrm{eff}(M) = \sum_{j} \Omega_j \int P_j(M,z) \frac{dV}{dz} dz 
\end{equation}
$\Omega_j$ is the solid angle of the good region in the field $j$ as described in Section \ref{sec:completeness}. $\frac{dV}{dz}$ is the differential comoving volume at redshift $z$. $M$ is the injected (absolute or apparent) magnitude bin. 

We use the $V_\textrm{max}$ method \citep{Schmidt1968,Efstathiou1988} to construct luminosity functions.
\begin{equation}
\begin{aligned}
    \phi(M) &= \frac{1}{\Delta M}\sum_{i=1}^{n}\frac{1}{V_{max,i}}\\
    &= \frac{1}{\Delta M}\sum_{i=1}^{n}\frac{1}{V_\textrm{eff}(M_i)}\\
\end{aligned}
\label{eq:vmax}
\end{equation}
The summation is over all galaxies whose magnitudes fall in the bin of interest. Because $V_{max}$ is the maximum volume that a galaxy can occupy while still being included the survey, we have to incorporate the completeness into the comoving volume calculation. $V_{max,i}$ is therefore an effective volume, $V_\textrm{eff}(M_i)$, interpolated at the magnitude $M_i$.

We employ a Monte Carlo approach to iteratively account for low-$z$ contaminants and to sample the intrinsic magnitudes of the galaxies. In each iteration, we model the low-$z$ contamination as a binomial process, where each galaxy candidate has a probability
$p_{\mathrm{contam}} = \lambda / N_{\mathrm{cand}}$
of being identified as a contaminant. Here, $\lambda$ is the expected number of interlopers calculated in Section~\ref{sec:contam}, and $N_{\mathrm{cand}}$ is the total number of candidates in the field. Candidates identified as contaminants are removed before further analysis. The intrinsic magnitudes of the remaining candidates are then drawn from the probability distribution functions (PDFs) derived in Section~\ref{sec:intrinsic_magnitude}. We then dynamically bin the remaining galaxies in magnitude space such that each bin contains at most 10 sources, while ensuring that the bin width does not exceed 0.5~mag. The left edge of the first bin is fixed to the nearest magnitude brighter than the brightest galaxy that is divisible by 0.25. This keeps the bin edges aligned to a fixed grid rather than the data itself, reducing potential bias from bin placement. At the end of each iteration, we compute the nominal luminosity function using Equation~\ref{eq:vmax}.  In Figure \ref{fig:LumfuncUV}, the green transparent data points show the individual luminosity functions from all iterations.

The final nominal luminosity functions are derived by re-binning the results from all iterations into uniform 0.5 mag intervals. Bins with no individual luminosity function measurements are treated as upper limits, corresponding to one galaxy per bin; these are shown as downward triangles in Figures \ref{fig:LumfuncUV} and \ref{fig:Lumfuncobs}, while measured bins are shown as squares. For bins with measured individual luminosity functions, the centers of the squares indicate the median values in both $\log_{10}\phi$ and magnitude. The uncertainties in the final nominal luminosity functions are computed as the quadrature sum of two components: the standard deviation across iterations and the Poisson noise, the latter estimated from the median number of galaxies per bin across all iterations. The horizontal error bars represent the full 0.5 mag width of each bin.

We then fit both Schechter and double-Schechter function to the final nominal luminosity functions. The fitting was performed by maximizing the total log-likelihood, which accounts for both detected data points and upper limits. For detected points, the likelihood was computed using the normal probability density function, while for upper limits we used the cumulative distribution function of a normal distribution. For each fit, we calculated the corrected Akaike information criterion (AICc), which is appropriate for model selection with small sample sizes. The single Schechter function is preferred in all cases based on the AICc. The best-fit Schechter functions are shown as solid lines in Figures \ref{fig:LumfuncUV}. Other observational results are plotted in blue, while theoretical predictions are shown in brown. Since theoretical predictions are often provided at integer redshifts, we linearly interpolated them in redshift space, where possible, to match the median redshifts of our sample.


\section{Results and Discussion}
\subsection{Rest-frame UV and B-band Luminosity Functions}
\subsubsection{Comparison with Previous Observations}
Numerous observational constraints for the UV luminosity function (LF) of galaxies at $z=5-9$ already exist in the literature. Here we present our measured UV LFs primarily as a sanity check and to facilitate comparison with our B-band LFs. Our results show good overall agreement with previous observational studies, though with some caveats. First, at $z\sim8$, our UV LF aligns more closely with recent JWST-based constraints and is higher than earlier HST- and ground-based measurements by 0.1-0.2 dex at the faint end \citep{Bowler2020,Bouwens2021,Leethochawalit2023} and up to 0.5 dex at the bright end. However, the bright end is in good agreement with HST-parallel survey results \citep[e.g.,][]{Leethochawalit2023,Rojas-Ruiz2024}. Second, our nominal UV LF at $z\sim5.2$ lies mostly between previously reported LFs at $z\sim5$ and $z\sim6$ results in literature. This offset is likely due to the stricter photometric redshift selection window we adopted ($z_p\in[4.5-6]$) compared to the sample in \citeauthor{Bouwens2021} ($4<z<6$).

As for the luminosity function (LF) in the rest-frame B band, prior to JWST, B-band luminosity functions were available only up to $z=5.5$ \citep[e.g.,][]{Giallongo2005,Gabasch2004,Poli2003}. The highest redshift constraints at $z=5.5$ are based on observations from the FORS Deep Field using ESO’s VLT telescopes, covering optical to near-infrared wavelengths up to the Ks band \citep{Gabasch2004,Heidt2003}. Our F435W dropouts ($z\sim5.2$) show consistency with these results within 1$\sigma$. We here show for the first time the B-band LFs at $z\sim7$ and 8 and extend the constraints at $z\sim5$ to a few magnitudes deeper.

\subsubsection{Comparison with Simulation Results}\label{sec:compareUVBwithlit}
To our knowledge, the only theoretical predictions for B-band luminosity functions at $z \gtrsim 6$ available in the literature are those presented by \citet{Ma2018}, shown as brown dashed lines in the right-hand panels of Figure~\ref{fig:LumfuncUV}. These predictions are derived from the \texttt{FIRE-2} simulations \citep{Hopkins2017}, which are cosmological zoom-in simulations with comoving volumes reaching up to $43^3 \ \mathrm{Mpc}^{3}$. The simulations employ the BPASSv2.0 stellar population synthesis models \citep{Eldridge2017}. 

It is important to note that \citet{Ma2018} provided B-band luminosity functions without accounting for dust attenuation, as dust-corrected luminosity functions are only available in the UV band. However, adopting the SMC bar attenuation curve—which has been demonstrated to closely approximate the dust attenuation of high-redshift galaxies \citep{Zafar2018}—the B-band attenuation is estimated to be roughly 3.5 times lower than that at 1500~\AA. Given that the maximum UV extinction in the \citeauthor{Ma2018} sample is 3~mag, this suggests that the B-band luminosity functions are much less unaffected by dust, unlike their UV counterparts.

We additionally extract B-band luminosity functions from the IllustrisTNG simulations \citep{Nelson2019}, using the TNG100 and TNG300 suites. These simulations have baryonic mass resolutions of $1.4 \times 10^{6} \ M_{\odot}$ and $1.1 \times 10^{7} \ M_{\odot}$, respectively, and are evolved in periodic boxes with comoving volumes of $110.7^{3} \ \mathrm{Mpc}^{3}$ and $302.6^{3} \ \mathrm{Mpc}^{3}$. We select these two most voluminous suites to ensure robust sampling of the bright end of the luminosity function. Although the mass resolutions differ by nearly an order of magnitude, the resulting distributions of subhalo physical properties—such as the stellar mass function—show only minor discrepancies (see \citealt{pillepich_et_al_2018smf}).

We determine B-band magnitudes for galaxies (i.e., subhalos, in the IllustrisTNG terminology) by summing the light from all stellar particles, without applying dust corrections. Specifically, we extract the \texttt{SubhaloStellarPhotometrics} field from the group catalogs, which provides photometry in eight bands: U, B, V, K, g, r, i, and z. After converting from Vega to AB magnitudes, we directly use the B-band values\footnote{The B-band magnitudes provided in IllustrisTNG are computed using the B3 response function from \citet{Buser1978}, which  resembles the Johnson–Cousins B filter in both wavelength coverage and overall shape. To be conservative, we also constructed the SDSS $g$-band LFs from the same TNG catalogs. The SDSS $g$ filter spans a similar wavelength range to the B-band but exhibits a right-skewed profile, whereas the B-band is left-skewed. The resulting $g$-band luminosity function agrees with the B-band one to within 0.05 dex, suggesting that the luminosity functions derived from the two B-band filters should be even more similar. This small difference is negligible compared to the variation observed between the TNG100 and TNG300 simulations.}. At $z=5$, our magnitude-limited samples ($M_B < -16$) contain $43{,}169$ galaxies in TNG100 and $658{,}121$ in TNG300.

Using the catalogs at $z = 5$, 6, 7, and 8, we bin galaxies by their B-band magnitudes in 0.5 mag intervals and compute number densities by dividing by the simulation volumes. We then fit Schechter functions to the resulting luminosity distributions. These fits, interpolated to match the redshifts of our observational samples, are shown as brown shaded regions in the right-hand panels of Figure~\ref{fig:LumfuncUV}, encompassing results from both TNG100 and TNG300.

When comparing our UV luminosity functions (LFs) with simulation predictions, we find that the LFs at lower redshifts are consistent within 1$\sigma$ of predictions from both the \texttt{FIRE-2} and \texttt{TNG} simulations \citep{Ma2018,Vogelsberger2020} when dust attenuation is included. However, at $z \sim 8$, our observed UV LF shows higher number densities than both simulations with dust, but it agrees better with the dust-free predictions from \citeauthor{Ma2018}. As discussed in \citet{Leethochawalit2023}, \citeauthor{Ma2018} may overestimate dust attenuation at high redshift, as they assume a local dust-to-metal ratio of 0.4 across the entire redshift range. However, there is growing evidence that the dust-to-metal ratio is strongly correlated with metallicity \citep{Wiseman2017,Li2019,Konstantopoulou2024}, and therefore likely decreases at higher redshifts, where metallicities are lower. \citeauthor{Vogelsberger2020} adopt a variety of dust models, resulting in a range of predicted LFs shown as the shaded brown region in Figure~\ref{fig:LumfuncUV}. These include models based on empirical UV slope--UV attenuation relations, as well as those incorporating redshift-dependent dust-to-metal ratios. However, all of their dust models are calibrated to match UV LFs from earlier HST observations at $z = 4$--$10$, so it is not surprising that they underestimate the UV LF at $z = 8$, as our results are higher than previous HST- and ground-based measurements.

Our measurements of B-band LFs partially agree with at least one theoretical prediction at each redshift, but no prediction agrees with our observations across all redshifts. The predictions from \citeauthor{Ma2018} agree well with our observations at higher redshifts but overpredict at $z\sim5$. Although \citeauthor{Ma2018} provide LFs in the B-band where dust was not included in the simulation, we do not expect dust to significantly affect the LF in the B-band. We therefore can reasonably conclude that our results are consistent with \citeauthor{Ma2018} at $z\sim7$ and $z\sim8$. However, there is some tension at $z\sim5$, where our B-band LF is up to 1 dex lower than their predictions at the bright end. The situation is reversed for the TNG results: our measurements agree better at $z\sim5$, especially in the fainter magnitude bins. TNG underpredicted the B-band LFs by up to 0.5 dex at $z\sim7-8$.

Identifying the root causes of the discrepancies mentioned above is challenging because cosmological hydrodynamical simulations are inherently complex and differ. A useful starting point is analytical models, where luminosity functions—currently still focused on UV LFs—are commonly used to constrain the star formation efficiency as a function of halo mass and redshift \citep[e.g.,][]{Furlanetto2017,Sipple2024,Dhandha2025}. The star formation efficiency parameter naturally encapsulates the combined effects of gas accretion and feedback. These models typically employ a constant conversion factor between star formation rate and luminosity, which is influenced by assumptions regarding the initial mass function (IMF), stellar population models, and dust attenuation. As a result, LFs are sensitive to numerous factors, including baryonic accretion rates, cooling rates, star formation density thresholds, feedback processes (e.g., radiative pressure, supernovae, stellar winds, and photoelectric heating), and the specific choices for IMF and stellar population models. These factors collectively influence the observed LFs, complicating direct comparisons between simulations and observations. 

\paragraph{Binary models}
If the discrepancies are attributed to the effect of binaries, our B-band LFs may suggest that binary models are necessary at $z\gtrsim7$. \citeauthor{Ma2018} use BPASSv2.0 stellar population synthesis models \citep{Eldridge2017} in their simulations, which include binary modeling of the stellar populations. The TNGs uses the single stellar population \citet[][BC03]{BC03} template library for star particles \citep{Nelson2015}, which does not include binary modeling. In the Milky Way and the Magellanic Clouds, the fraction of binary and multiple star systems is estimated to be around 40\% among F, G, and K-type stars, while it may reach 80\% for O and B-type stars  \citep{Sana2014,Frost2025}. Moreover, the multiplicity of stars may increase with decreasing metallicity \citep{Bate2019}, suggesting that higher redshifts with lower metallicity may have even higher binary fractions. These need to include binary systems especially at high redshift may explain why the measured B-band LFs at $z\gtrsim7$ agree better with simulations with binary stellar populations. 

On the other hand, our $z\sim5$ B-band LF result (the bottom right panel of Figure \ref{fig:LumfuncUV}) is 0.5 to 1 dex smaller than the predictions by \citeauthor{Ma2018}, but it agrees better with the TNG results. If binary models are effective at higher redshifts, why do they fail at our lowest redshift bin? One possible explanation is that the binary fraction in the BPASS model does not vary with star types. As noted by \citet{Eldridge2017}, BPASS was initially optimized for young populations, with poorly constrained binary fractions for lower-mass stars (a few solar masses and below). At $z\sim5$, the universe is approximately 1.2 Gyr old, which is old enough for some B- and A-type stars to evolve into Asymptotic Giant Branch (AGB) stars where binary interactions become significant. An overestimated interacting binary fraction in low-mass stars could therefore contribute to the discrepancy between \citet{Ma2018} and observations.

\paragraph{Stellar Population Synthesis Codes for Single Stars} The differences in the predicted LFs between the two simulations at $z\sim5$ may also arise from the choice of stellar population synthesis codes for single stars. \citet{Eldridge2017} compared their BPASS single-star models—used as the basis for their binary models—to Starburst99, BC03, and \citet{Maraston2005} models. They found that the stellar populations older than a few hundred Myr in the BPASS models are significantly redder than those in the BC03 models. Since some stellar populations can reach several hundred Myr at $z\sim5$, using a redder model like BPASS could consequently lead the B-band LFs in \citet{Ma2018} to have larger number density than the TNG results.

\paragraph{IMFs} We do not consider differences in the initial mass functions (IMFs) to be a significant factor affecting the differences in the luminosity functions (LFs). This is because the IMFs used in the two simulations—Kroupa (\citeyear{Kroupa2001}) in \citeauthor{Ma2018} and Chabrier (\citeyear{Chabrier2003}) in the TNGs—are identical for stars above one solar mass. Since stars with masses below one solar mass do not significantly contribute to the light in either the UV or B-band at $z \gtrsim 5$, their impact is negligible. Although some spectral observations of galaxies at $z > 5$ hint at a more top-heavy IMF in the early universe \citep[e.g.,][]{Cameron2024}, which could enhance UV emission \citep{Jeong2025}, \citet{Cueto2024} argued that a top-heavy IMF alone cannot explain the high number density of UV LFs at these redshifts due to the strong stellar feedback it generates.

\paragraph{Extreme Emission Lines}  
We now consider whether extreme emission line galaxies (EELGs) could contribute to the high number densities observed in the B-band luminosity functions (LFs) at $z \gtrsim 6.4$. EELGs are galaxies undergoing intense bursts of star formation, characterized by large rest-frame equivalent widths (EWs; $>100$\,\AA) in [OIII]$\lambda5007$ and H$\alpha$ emission lines. In the sample of star-forming galaxies from \citet{Boyett2024}, 75\% of galaxies at $z>5.7$ exhibit [OIII]$\lambda5007$ EWs exceeding 500\,\AA, with some reaching up to 3000\,\AA. This fraction drops to 40\% for galaxies at $3<z<5.7$. The elevated incidence of EELGs at $z>5.7$ is consistent with the high number densities we observe in the B-band LFs from our F606W and F090W dropout samples. Since theoretical predictions from \citet{Ma2018} and TNG simulations in Figure \ref{fig:LumfuncUV} are based on stellar particles and do not include emission lines, this raises the possibility that the observed excess relative to the TNG predictions arises not from stellar population or binary evolution models, but from emission line contributions.

However, while EELGs may contribute to the observed excess, they are unlikely to fully account for the discrepancy. For a filter with a $\sim$1000\,\AA\ bandwidth and a top-hat transmission profile, rest-frame EWs of 500--3000\,\AA\ can boost the observed magnitude by approximately 0.5--1.2 mag. In the case of the B-band, however, [OIII]$\lambda5007$ falls within a region where the filter response is only about 25\% of the peak transmission. This leads to a smaller flux boost effectively shifting the observational LFs to the left by approximately 0.1--0.6 mag. In practice, the boost is likely toward the lower end (a few tenths of a magnitude), as \citet{Boyett2024} report that most galaxies exhibit [OIII] EWs in the 500--1000\,\AA\ range, with fewer reaching above 1000\,\AA. However, the observed B-band LFs at $z \gtrsim 6.4$ are shifted by approximately 1 mag to the left relative to TNG predictions—larger than what can be explained by emission line boosting alone. We also note that we do not expect significant contributions from [OII]$\lambda3727$, as EELGs typically show [OIII]/[OII] ratios greater than 10—much higher than those observed in local galaxies \citep{Cameron2023}. We will explore the role of EELGs further in the apparent luminosity function section. 

\paragraph{AGNs} Other factors that have been explored in the literature to affect luminosity functions include AGN feedback. Large-area surveys such as the Subaru/Hyper Suprime-Cam and the CFHT Large-Area U-band surveys have demonstrated that when point sources are not excluded (i.e., AGNs are included), the UV luminosity functions (LFs) at $z=2$–$7$ exhibit a bright-end excess relative to a single Schechter function. In such cases, the data are better described by combinations of double power-law (DPL+DPL) or DPL+Schechter functions \citep[e.g.,][]{Harikane2022,Finkelstein2022,Adams2023}. To explain this, \citet{Trinca2024} showed that their semi-analytical model, which includes luminosity contributions from accreting black holes, successfully reproduces the observed UV LFs at $z=4$–9. In their model, AGN light dominates the bright end, while stellar populations dominate the faint end, with a transition near $M_{\mathrm{UV}}\sim-23$ where both contributions become comparable.

In Section \ref{sec:LFmethod}, we showed that single Schechter functions provide good fits to all of our luminosity functions. For the UV, this is expected because our UV LFs only extend to magnitudes where the galaxy population is expected to dominate, i.e., to $M_{\mathrm{UV}}\sim -22.5$ for the F090W and F435W dropouts and to $M_{\mathrm{UV}}\sim -22$ for the F606W dropouts. According to \citet{Trinca2024}, these magnitude limits correspond to AGN contributions of roughly 30\% in UV luminosity for the F090W and F435W dropouts and less than 10\% for the F606W dropouts—small enough that a single Schechter form remains adequate.

In the B-band LFs, however, we suspect a larger AGN contribution. The brightest bins of the F090W and F435W dropouts appear slightly elevated relative to the best-fit Schechter models, though still consistent within $2\sigma$. This mild excess could indicate AGN contamination. \citet{Trinca2024} adopted $\frac{L_{\mathrm{bol}}}{L_B} = 5.13$ and $L_\nu \propto \nu^{-0.44}$ to convert AGN bolometric luminosities ($L_{\mathrm{bol}}$) to B-band luminosities ($L_B$) and subsequently to UV luminosities. Under these assumptions, AGNs should be roughly twice as luminous in the B-band as in the UV when expressed in terms of per-frequency luminosity $L_\nu$, which is the quantity directly corresponding to AB magnitudes. Consequently, an AGN contribution of $\sim$30\% in the UV could rise to $\sim$60\% at the bright end of the B-band LF. We note, however, that these apparent “kicks” in the B-band LFs could also be contributed by other effects discussed earlier, such as strong emission lines. We encourage future theoretical models that include AGN contributions to extend their predictions to additional bands, and for observational efforts to cover wider areas to determine whether these bright-end excesses are genuine.

For completeness, we highlight that the only study to our knowledge that directly compares AGN effects on luminosity functions in non-UV bands is \citet{TaylorKobayashi2016}. However, their cosmological chemodynamical simulations are limited to $z=0-5$. They found that AGN feedback decreases the number of bright galaxies and increases the number of faint galaxies by up to $\sim$0.5 dex at $z=5$, opposite to the trend predicted by \citet{Trinca2024} at higher redshifts. This behavior arises because AGNs in their model suppress star formation in massive galaxies while delaying early star formation in low-mass galaxies, weakening supernova feedback and preventing quenching. However, both their AGN and non-AGN simulations overpredict the observed UV and B-band LFs at $z\sim3$ by 1–2 dex, and their B-band LFs also exceed ours at $z\sim5$, limiting direct comparison.

\citet{Cowley2018} explored the impact of star-formation feedback on UV luminosity functions but found that the effect of dust dominates over that of feedback in the ultraviolet. Specifically, they examined two recipes for supernova feedback in their semi-analytical models: one using a mass-loading factor dependent only on galaxy circular velocity, and the other incorporating an additional redshift dependence. Their findings revealed that, without dust, the UV LFs predicted by the evolving feedback recipe is higher than the fixed recipe by as large as 1 dex at all UV magnitudes. However, when dust is included, the difference remains only at the faint end of the UV LF ($M_\mathrm{UV} > -19$), fainter than our results here. Given that dust has a lesser effect on redder bands, we hypothesize that the impact of feedback variations should be more apparent in the B-band LF. We encourage future theoretical work to further explore this possibility.

Lastly, we do not believe that our luminosity function (LF) measurements are significantly affected by the selection criteria. Traditionally, high-redshift galaxies have been identified using Lyman-break galaxy (LBG) selection techniques \citep[e.g.,][]{Oesch2007,Bouwens2011,Trenti2011}. More recent studies have increasingly relied on photometric redshifts, either in combination with LBG color selection \citep[e.g.,][]{Bouwens2015,Roberts-Borsani2022,Morishita2023} or as standalone criteria \citep[e.g.,][]{Finkelstein2015,Adams2024,Hainline2024}. Our selection approach is in between photometric selection and LBG selection. Specifically, it requires detection in both the rest-frame UV band and the band of interest. This two-band detection criterion is comparable to that of \citet{Finkelstein2015}, ensuring the exclusion of spurious sources. While requiring detection in the rest-frame UV may bias against selecting quiescent galaxies, the quiescent fraction at $5<z<7$ is reported to be below 10\% based on the specific star formation rate (sSFR) criteria in the highest stellar mass bin \citep{Russell2024}. This fraction is even lower when using the UVJ diagram method, which is more relevant to our analysis. The fraction of 10\% corresponds to less than 0.05 dex in number density in log scale, which is minor compared to other sources of uncertainty. Moreover, our LF measurements at $z\sim5$ are consistent with those of \citet{Gabasch2004}, who employed photometric redshift selection. Additionally, \citet{Liu2016} predicted UV LFs at $z\geq5$ using both all sample from their semi-analytic/semi-numerical models and those selected with LBG selection techniques, finding no significant difference in the results. Finally, any selection effects are further accounted for in our completeness simulations, under the assumption that the injected SEDs adequately represent the population's range.

\begin{figure*}
    \centering 
    \includegraphics[width=0.45\textwidth]{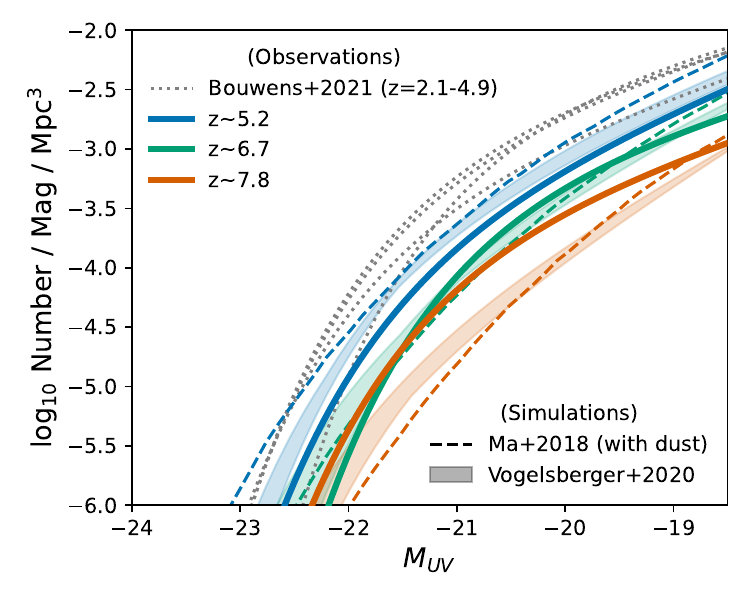}
    \includegraphics[width=0.45\textwidth]{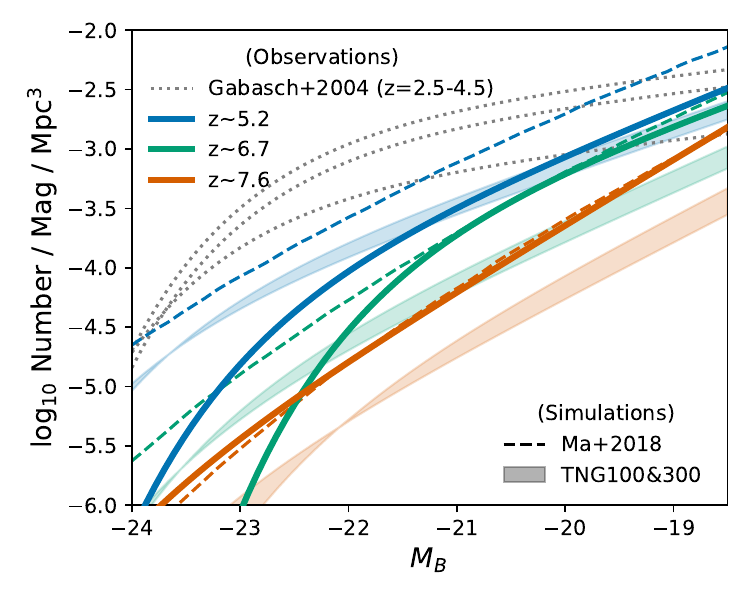}
    \caption{Evolution of the luminosity functions in the rest-frame UV (left panel) and B band (right panel) for three dropout samples: F435W dropouts ($z \sim 5.2$, blue), F606W dropouts ($z \sim 6.7$, green), and F090W dropouts ($z \sim 7.8$, red). Observational constraints at lower redshifts from \citet{Bouwens2021} and \citet{Gabasch2004} are shown as dotted gray lines. Simulation results from \citeauthor{Ma2018} and the TNG simulations, interpolated to the corresponding redshifts, are shown as dashed and shaded lines in matching colors.}
    \label{fig:redshiftevol}
\end{figure*}

\subsubsection{Redshift Evolution}
We replot the best-fit Schechter functions for the rest-frame UV and B-band luminosity functions in Figure~\ref{fig:redshiftevol} to better illustrate their redshift evolution. Compared with observational results at lower redshifts (gray dotted lines), the evolution in the B-band appears more pronounced than in the UV. In particular, our results indicate a rapid decline in the B-band number density between $z \sim 4$ (the lowest gray dotted line) and $z \sim 5$ (the blue solid line). However, we are unable to draw a firm conclusion regarding the bright-end behavior of the B-band LF for the F606W dropouts ($z \sim 6.7$)—specifically, whether it falls below that of the F090W dropouts—as our nominal luminosity functions provide only upper limits beyond $M_B = -22$ mag. Another notable feature is that the inferred faint-end slopes of our B-band luminosity functions are significantly steeper than those reported by \citet{Gabasch2004}, but are broadly consistent with the results from \citet{Ma2018} and the TNG simulations. This suggests a continued rise in number density toward fainter magnitudes.

When compared with simulations over the same redshift range of $z=5.2-7.8$ (dashed and shaded lines), the observed luminosity functions show a more moderate evolution—roughly half the strength predicted by simulations in both the UV and B bands. Both the FIRE and TNG simulations predict a comparable evolution of approximately $\sim$1 dex. The gradual evolution found here extend the smooth bright-end evolution of the UV luminosity functions, reported in previous studies \citep[e.g.,][]{McLure2013, McLeod2015, Finkelstein2016, Leethochawalit2023, Rojas-Ruiz2024, Robertson2024}, to the B-band.

\begin{figure*}
    \centering 
     \includegraphics[width=0.46\textwidth]{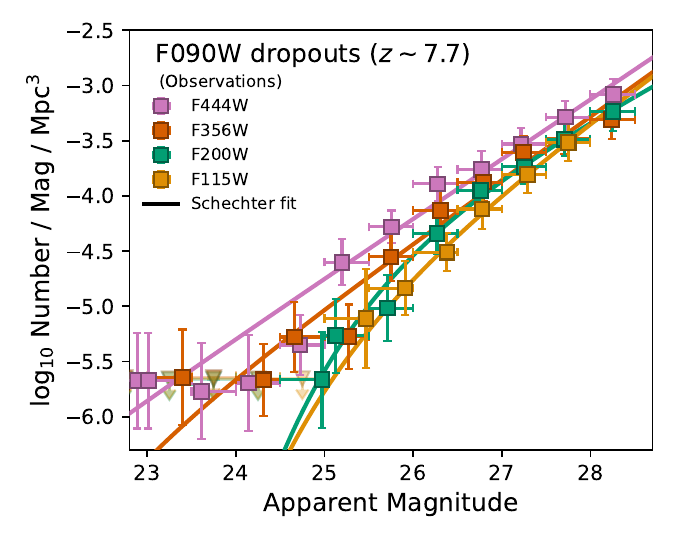}
    \includegraphics[width=0.46\textwidth]{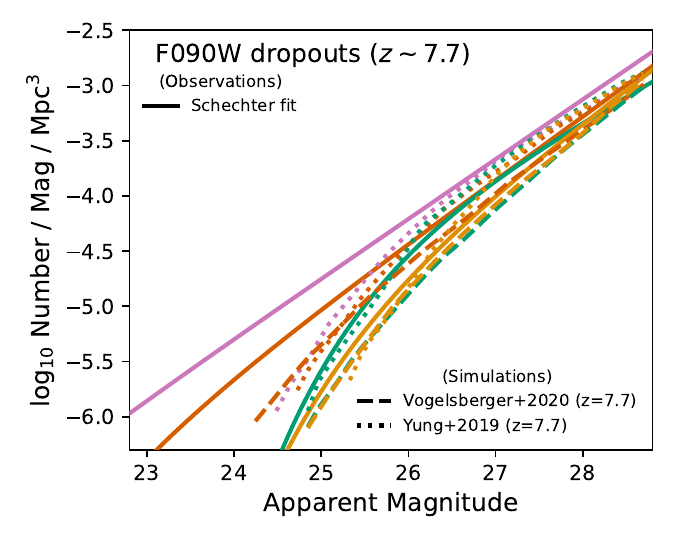}\\
    \includegraphics[width=0.46\textwidth]{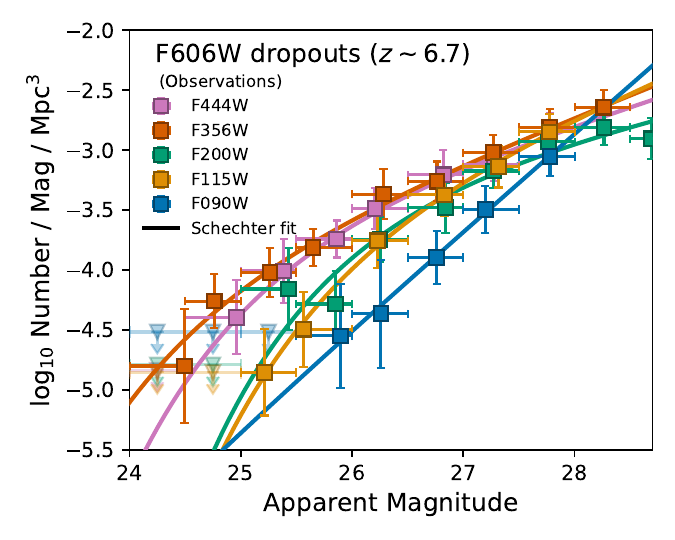}
    \includegraphics[width=0.46\textwidth]{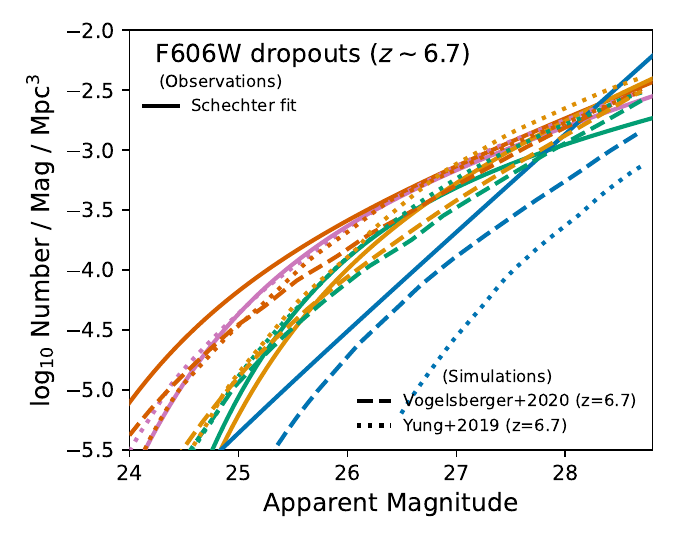}\\
    \includegraphics[width=0.46\textwidth]{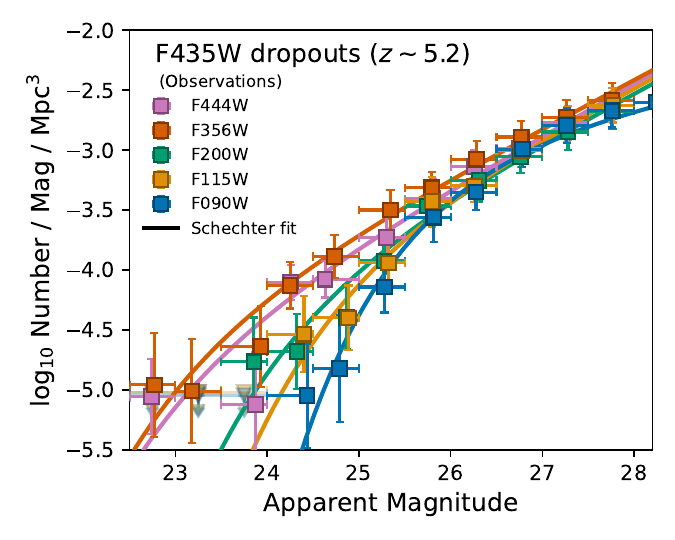}
    \includegraphics[width=0.46\textwidth]{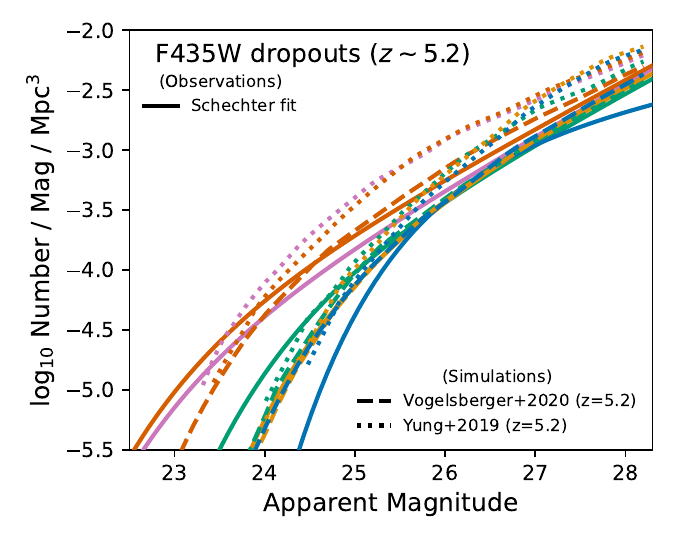}\\
    \caption{(Left) Apparent luminosity functions in various NIRCam/JWST filters—F444W (pink), F356W (red), F200W (green), F115W (yellow), and F090W (blue)—shown for three redshift bins indicated in the top-left corner of each row. Square symbols represent the measured luminosity functions, and solid lines show the best-fit Schechter functions. Upper limits are plotted in lighter shade. (Right) Comparison of the best-fit Schechter functions (solid lines) with simulation results from \citet{Vogelsberger2020} and \citet{Yung2019}, plotted as dashed and dotted lines in corresponding colors. The nominal luminosity functions and the corresponding best-fit Schechter parameters shown in this figure are also provided as the Data behind the Figure.}
    \label{fig:Lumfuncobs}
\end{figure*}
\subsection{Apparent JWST band Luminosity Functions}
The advantage of using apparent luminosity functions is that they are less affected by uncertainties and biases in photometric redshift estimates, since they do not require luminosity distances to convert the apparent magnitudes to absolute magnitudes, provided that galaxies are assigned to the correct redshift bin. In Figure~\ref{fig:Lumfuncobs}, we present the apparent luminosity functions in the JWST/NIRCam F090W, F115W, F200W, F356W, and F444W bands. The predictions from literature are also linearly interpolated from integer redshifts to match the median redshift of our sample. 

In the highest redshift bin (upper panels), the best-fit Schechter functions in the two reddest bands, F356W and F444W, appear to lie above the predictions from \citet{Vogelsberger2020} and \citet{Yung2019} from $m < 26$ to the brightest magnitudes. 

At $z \sim 6.7$ (middle panels), our best-fit luminosity functions are generally consistent with the predictions from both \citet{Vogelsberger2020} and \citet{Yung2019} across most magnitudes. The exception is the bluest F090W band, where a discrepancy is observed. However, this difference is difficult to interpret, as the F090W band partially overlaps with the blue side of the Lyman dropouts, and the offset may result from the redshift interpolation used for the simulation predictions.

At $z \sim 5$ (lower panels), the observed luminosity functions at the faint end in the two reddest bands are systematically lower than the predictions from \citet{Yung2019} but are in overall agreement with the TNG predictions from \citet{Vogelsberger2020}. However, the bright-end show a possible excess relative to both simulations. At $m \sim 23$, the observed number densities may exceed the TNG predictions by up to 1 dex at the $2\sigma$ level. We note that these two bands are redder than F277W, which corresponds to the observed band for the rest-frame B-band. 

Similar to the discussion in Section~\ref{sec:compareUVBwithlit}, we find that extreme emission line galaxies cannot fully account for the tentative excess observed in the apparent luminosity functions in the F356W and F444W bands relative to the TNG predictions. At $z \sim 5.2$, the [OIII]$\lambda5007$ emission line falls within the F356W filter, while the H$\alpha$ line lies within the F444W filter. The widths of both F356W and F444W are approximately 1500\AA\ in rest-frame for $z \sim 5$ galaxies. Assuming the rest-frame filter width of 1500\AA, an emission line with an EW of 1000\AA\ could enhance the flux within the filter by approximately at most 0.5 mag, not enough to explain with the observed bright-end excess in the F356W and F444W LFs relative to the TNG predictions for F435W dropouts. For F090W dropouts, the [OIII]$\lambda5007$ line falls within the F444W filter, while the F356W filter is not expected to contain any strong emission lines. Yet, we observe an excess in both F444W and F356W, indicating that EELGs alone cannot explain the excess seen at $z \sim 8$.

The excess seen in red observed bands might share a common origin with the B-band bright-end excess behavior, potentially reflecting AGN activity and/or a milder manifestation of the newly identified red population at high redshifts—the extremely red objects (EROs). EROs, sometimes referred to as ``little red dots'', are characterized by red rest-frame optical colors while retaining blue or flat slopes in the rest-frame UV \citep[e.g.,][]{Labbe2023,Furtak2023}. These objects are likely either dusty star-forming galaxies or obscured AGNs \citep{Barro2024}. While none of our F356W-selected and F444W-selected candidates at $z \sim 7$ and $z \sim 8$ display the extreme red colors observed in EROs (F277W$-$F444W $> 1.5$ mag), approximately 20 candidates at each redshift do exhibit somewhat red colors (F277W$-$F444W $> 0.5$ at $z \sim 7$ and F277W$-$F444W $> 0$ at $z \sim 8$), along with flat UV slopes ($|$F150W$-$F200W$| < 0.5$). This trend may point to a contribution from dusty galaxies or obscured AGNs—broadly consistent with the AGN-driven interpretation considered for the B-band bright-end excess.
\begin{figure*}
    \centering 
    \includegraphics[width=0.9\textwidth]{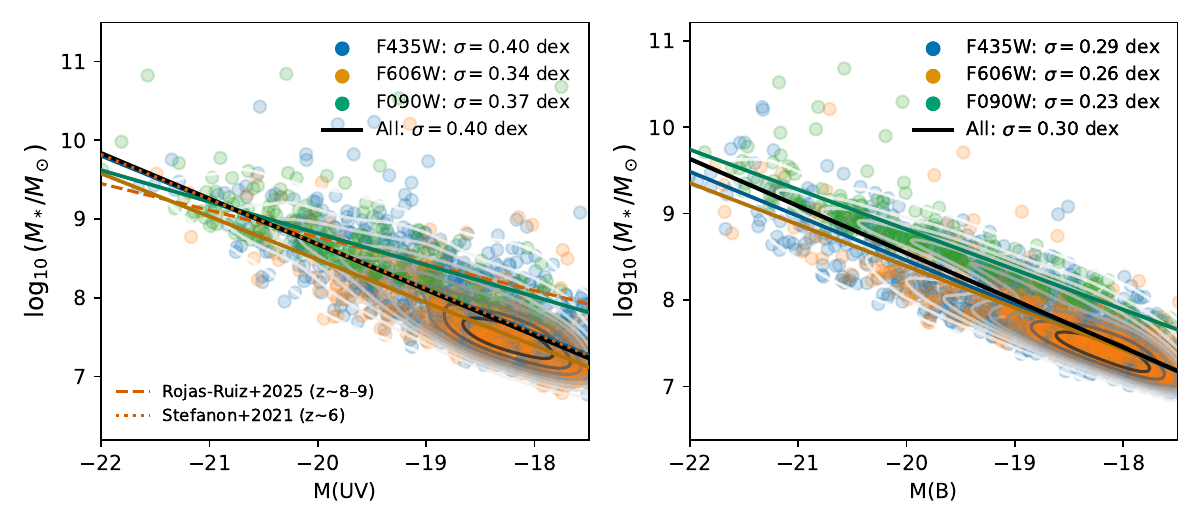}
    \caption{M/L ratio relations in the rest-frame UV (left) and B band (right). Data points are color-coded by redshift sample, with corresponding best-fit linear relations shown as solid lines. The best-fits relation for all sample (black lines) can be described as $\log_{10}(M_*/M_\odot)=-0.58(M_{UV}+20) + 8.68$ and $\log_{10}(M_*/M_\odot) = -0.55(M_B+20)+8.54$. The labeled $\sigma$ values indicate the scatter in each relation. For comparison, reference relations from previous studies \citep{Stefanon2021, Rojas-Ruiz2024} are overplotted as red dotted and dashed lines.}
    \label{fig:masstolight}
\end{figure*}

\subsection{Rest-frame Luminosity - Stellar Mass relations}
The stellar mass-to-light (M/L) ratio is a scaling relation that links an observable quantity (luminosity) to a physical property (stellar mass), providing a means to connect luminosity functions (LFs) to stellar mass functions \citep[e.g.,][]{Gonzalez2011}, as well as a diagnostic of galaxy formation histories \citep[e.g.,][]{Stark2009}. This relation is typically expressed as a log-log correlation between absolute magnitude and stellar mass. Most previous studies have focused on the UV, finding that a linear relation offers a reasonable description of galaxies at $z > 4$ \citep[e.g.,][]{Duncan2014,Song2016,Bhatawdekar2019}. Even with the data from JWST, recent results show that the scatter in the $M_{\mathrm{UV}}$--$M_*$ relation remains quite large, with $\sigma \sim 0.45$ \citep{Rojas-Ruiz2024,Morishita2024}.

We find that the M/L ratio relation for B band is tighter than the M/L ratio relation for the UV band. The plots for the two relations are shown in Figure \ref{fig:masstolight}. We show the best-fit relations for both individual redshift samples and the combined sample. In the UV band, we find the scatter of 0.4 dex, similar to what was found in previous studies. In all cases, the M/L ratio relations in the B-band are approximately 0.1 dex tighter.

The tighter M/L ratio relation in the B-band is expected, yet it underscores the importance of rest-frame optical LFs in linking observables to the underlying physical properties of galaxy stellar populations. UV light is strongly influenced by recent star formation history and does not necessarily correlate with the total stellar mass \citep{Wilkins2013}. Specifically, UV light at 1500~\AA\ is sensitive to stellar populations with ages of $\sim10^7$ years--comparable to typical starburst timescales--while light at wavelengths longer than 3000~\AA\ traces longer-lived stars (A-type and later), which dominate the stellar mass budget \citep{Conroy2013}. Moreover, UV light is particularly sensitive to dust attenuation, further reducing its reliability as a tracer of stellar mass \citep{Wu2020}. These factors make rest-frame optical LFs a more robust probe of stellar mass buildup, and perhaps overall stellar population properties of the galaxies.

We note that the scatter reported here is calculated from the stellar masses measured by \texttt{Bagpipes} described in Section \ref{sec:absmag_measurement}. Although stellar mass is generally the most reliable physical parameter derived from SED fitting, it is sensitive to the assumed SFH, the wavelength coverage of the data used, and the adopted SPS model \citep{Conroy2013}. The current M/L ratio relation in the UV band derived by different studies, which differ in wavelength coverage, number of filters, and SED fitting models, varies by as large as 1 dex at $z=8-9$ \citep{Rojas-Ruiz2024,Stefanon2021}. We therefore refrain from comparing the numerical values of our relation across the redshift range (as they come from different data sets with different wavelength coverage in rest frame). We defer further investigation in the M/L ratio beyond the scatter amount as well as the derivation of stellar mass function to future studies.

\section{Conclusion} \label{sec:conclusion}
We have presented new measurements of galaxy luminosity functions (LFs) across rest-frame and observed-frame bands at redshifts $z\sim5$–9, using deep photometric data from JWST/NIRCam. Specifically, we constructed rest-frame UV and B-band LFs—extending the rest-frame B-band constraints to $z\sim7$ and 8 for the first time—as well as apparent LFs in F090W, F115W, F200W, and F356W filters. Our sample comprises 1374 F435W-dropouts ($z\sim5.2$), 669 F606W-dropouts ($z\sim6.7$), and 405 F090W-dropouts ($z\sim7.8$), selected using consistent photometric redshift and dropout criteria.

To ensure accurate completeness corrections and photometric bias correction, we employed injection-recovery simulations with the \texttt{GLACiAR2} framework. Rest-frame absolute magnitudes were measured using \texttt{eazy}, with validation from SED fits using \texttt{Bagpipes}. Luminosity functions were derived via the $1/V_\mathrm{max}$ method with Monte Carlo sampling to account for flux uncertainties and the Eddington bias.

Our main findings are:
\begin{itemize}
    \item{Our rest-frame UV LFs are in good agreement with existing JWST-based measurements. For the first time, we present B-band luminosity functions at $z\sim7$ and $z\sim8$, and extend previous results at $z\sim5$ to $M_B=-18$.}
    \item{No single simulation reproduces all observed trends across the UV and B-band LFs. While our B-band LF at $z\sim5$ aligns well with TNG predictions, those at higher redshifts are more consistent with the FIRE-2 simulations. This redshift-dependent agreement suggests a potential change in stellar population properties, including the role of binaries and the choice of stellar population synthesis models.}
    \item{The contribution of EELGs—galaxies with high-equivalent-width [OIII]+H$\beta$ lines—is likely non-negligible at $z \gtrsim 6$, and may explain part of the excess seen in rest-frame B, F356W and F444W LFs. However, line boosting alone likely cannot account for the full discrepancy with simulation predictions, especially given filter response profiles and typical observed equivalent widths.}
    \item{A mild bright-end excess is observed in the B-band luminosity functions, though not statistically significant, and may hint at a contribution from AGNs.}
    \item{B-band LFs evolve with redshift more strongly than their UV counterparts. However, both UV and B-band LFs exhibit a smaller redshift evolution than predicted by simulations. While FIRE and TNG simulations forecast a $\sim$1 dex decline from $z\sim5$ to 8, our observations indicate a more modest decline at half a dex, implying that massive star-forming galaxies remain more common at redshifts considered (up to $z=9$).} 
    \item{Our luminosity functions in NIRCam bands broadly match predictions at $z\sim5$. However, apparent luminosity functions in F356W and F444W show possible bright-end excess relative to simulations. While extreme emission line galaxies (EELGs) may partially account for this excess, the observed trends may also suggest the presence of moderately red sources—possibly dusty star-forming galaxies or obscured AGNs—resembling a milder version of the recently reported “little red dots.”}
    \item{The rest-frame B-band shows a tighter correlation with stellar mass than the UV. This highlights the value of B-band luminosity functions in constraining stellar mass functions and interpreting the stellar population properties of galaxies, even at early times.}
\end{itemize}

These results provide new empirical benchmarks for constraining models of early galaxy formation, star formation efficiency, and dust attenuation. Future deeper and wider JWST observations will further refine these constraints, especially in the rest-frame optical regime.

\begin{acknowledgments}
This work is supported by the Fundamental Fund of Thailand Science Research and Innovation (TSRI) through the National Astronomical Research Institute of Thailand (Public Organization) (FFB680072/0269). T.W. is supported by SUT, TSRI, and NSRF (grant number 195242). The authors acknowledge the use of the Chalawan High Performance Computing cluster, operated and maintained by the National Astronomical Research Institute of Thailand (NARIT).
\end{acknowledgments}

%

\vspace{5mm}
\facilities{HST(STIS), JWST(STIS)}




\bibliography{sample631}{}
\bibliographystyle{aasjournal}



\end{document}